\documentclass[aps,prl,twocolumn,showpacs,amsfonts,a4]{revtex4}
	
\usepackage[dvips]{graphics}
\usepackage{shortvrb}
\MakeShortVerb{\!} 
\newcommand{\Z}{\mathbb{Z}_2}
\newcommand{\re}{\left|D_0\right>}

\begin{document}

\title{
Quantum Dimer Model on the  Kagome Lattice: Solvable Dimer Liquid and
Ising Gauge Theory}

\author{G.~Misguich}
\email{gmisguich@cea.fr}

\author{D.~Serban}

\author{V.~Pasquier}

\affiliation{
Service de Physique Th\'eorique, 
CEA/Saclay
91191 Gif-sur-Yvette c\'edex, France}

\bibliographystyle{prsty}

 
\begin{abstract}

We introduce quantum dimer  models on lattices made of  corner-sharing
triangles.   These  lattices includes the  kagome   lattice and can be
defined  in  arbitrary geometry.  They   realize fully  disordered and
gapped dimer-liquid  phase with  topological degeneracy and deconfined
fractional  excitations, as well   as solid phases.  Using geometrical
properties of the  lattice,  several  results are obtained    exactly,
including  the full spectrum of a  dimer-liquid.  These models offer a
very    natural -  and maybe  the   simplest  possible  - framework to
illustrate general concepts   such as  fractionalization,  topological
order and relation to $\mathbb{Z}_2$ gauge theories.

\end{abstract}
\maketitle

%_______________________________________________________________________________

Quantum dimer  models     (QDM)   were introduced    by  Rokhsar   and
Kivelson~\cite{rk88} in the context  of resonating valence-bond  (RVB)
theories for the  high-temperature  superconductors~\cite{anderson87}.
Such   models are expected to  describe  the dynamics of singlet bonds
(dimer) in   quantum   disordered spin-$\frac{1}{2}$ antiferromagnets.
They can describe two  generic  phases: spin-liquids where  the system
breaks no  symmetry at all  and dimer (or valence-bond) crystals where
long-range  dimer-dimer   correlations develops.    Recently a genuine
liquid  phase with a finite  correlation length was  found in a QDM by
Moessner and Sondhi   on  the triangular  lattice~\cite{ms01}.    Such
liquid states  have attracted a lot  of interest because  they display
both fractional excitations and topological order~\cite{wen91}.  While
fractionalization could play  an  important role  in  some theories of
high-temperature      superconductors~\cite{anderson87,sf01},      the
topological properties  of these liquid  states have  been proposed as
possible  devices   to    implement    quantum   bits    for   quantum
computations~\cite{kitaev,ioffe02}.

In this  Letter  we introduce  QDM which  realize such a  dimer-liquid
phase.   Due  to   the  geometric properties    of   lattices made  of
corner-sharing triangles  (the simplest  two-dimensional example being
the kagome  lattice~\footnote{The QDM discussed in   this work {\em do
not} aim at describing   the physics of the spin-$\frac{1}{2}$  kagome
antiferromagnet.  We are currently  investigating a different  QDM for
this later  problem.}), these models are solved  exactly.  As  for the
solvable     point  of  the   QDM   on    the  square~\cite{rk88}  and
triangular~\cite{ms01}   lattices  (see      also    Refs.~\cite{ns01}
and~\cite{bfg01}),     the  ground-state    is     the equal-amplitude
superposition   of  all dimer    coverings   in a   given  topological
sector~\footnote{Dimer coverings are grouped in topological sectors by
considering their transition graphs, that is the set of loops obtained
by superposing two coverings on top of each other.  Two coverings with
only topologically trivial loops are in the same sector.}.  Such state
has been first considered  by Sutherland~\cite{s88} (SRK state in  the
following) and is    the prototype of    resonating valence-bond (RVB)
state.    However, our model  has   several important differences with
previous analogs~: 1) Not only the ground state but all excited states
wave  functions are known.  Elementary  excitations are (pairs of) non
interacting and gapped vortices~\cite{rc89k89} (called {\em visons} in
the recent literature~\cite{sf01}).  2) The model can be solved on any
geometry: torus,   discs or spheres.   This  allows to investigate the
interplay   between  topology, ground-state  degeneracy and elementary
excitations in an very  simple  way.  3) Dimer-dimer correlations  are
strictly  zero  above one lattice spacing.   This  makes this SRK wave
function  the most  possible disordered  dimer liquid  state.  4) This
state is {\em inside} the liquid  phase, it is not  lying at the phase
boundary with  a crystalline phase.  5)  It is known~\cite{msf02} that
QDM  can be obtained   as special limits  of  $\Z$ gauge theories, the
gauge  variable  being the dimer   number on a bond.    Here we show a
complete  equivalence with   a  $\Z$ gauge   theory.  This allows   to
investigate   the confinement  transition   which goes  with  a  dimer
crystallization  in a  simple QDM  which  exhibits a  liquid to  solid
transition accompanied by a vison condensation.

For all these reasons the model we introduce is  more than a toy model
but the simplest possible  RVB liquid. It  is a ``free dimer  liquid''
point in the short-ranged RVB  phase. It shows the generic  properties
which characterize that state of matter but almost any quantity can be
computed exaclty.  It is  a  natural starting  point  for perturbative
expansions toward more realistic models.

\begin{figure}
	\begin{center}
	\resizebox{!}{1.25cm}{\includegraphics{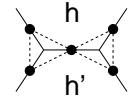}}
	\vspace*{-0.3cm}
	\end{center}
	\caption[99]{Medial  lattice construction.    From a trivalent
	lattice $H$ (full lines)    we construct a lattice $K$   whose
	sites (black dots) are centers of the bonds.  The sites of $K$
	are  linked  together   (dashed   lines) to   form  triangular
	plaquettes.  The pseudospin variables  live on the dual of $H$
	(h and h').}\label{medial}
\end{figure}

\begin{figure}
	\begin{center}
	\resizebox{2.5cm}{3.2cm}{\includegraphics{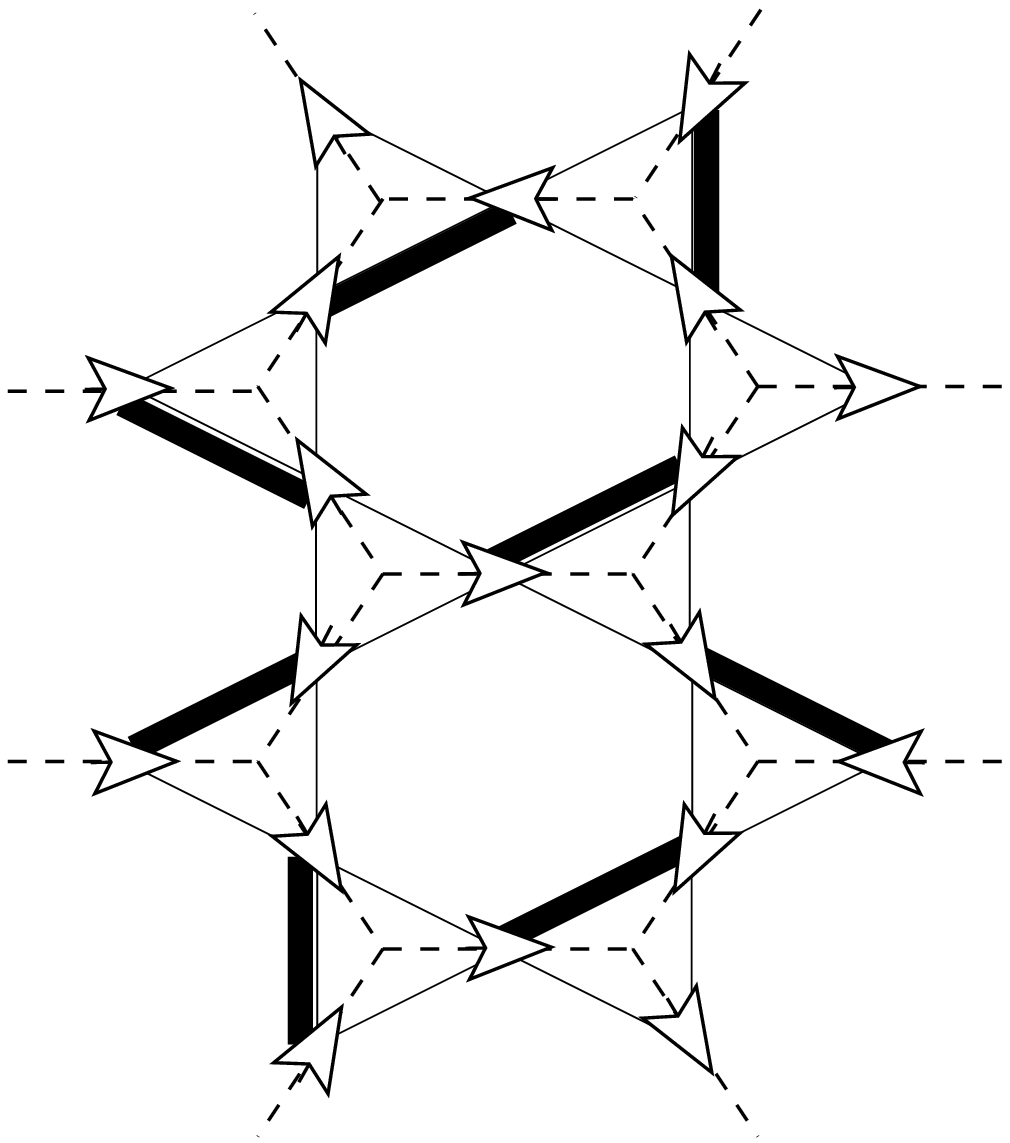}}
	\resizebox{3.9cm}{3cm}{\includegraphics{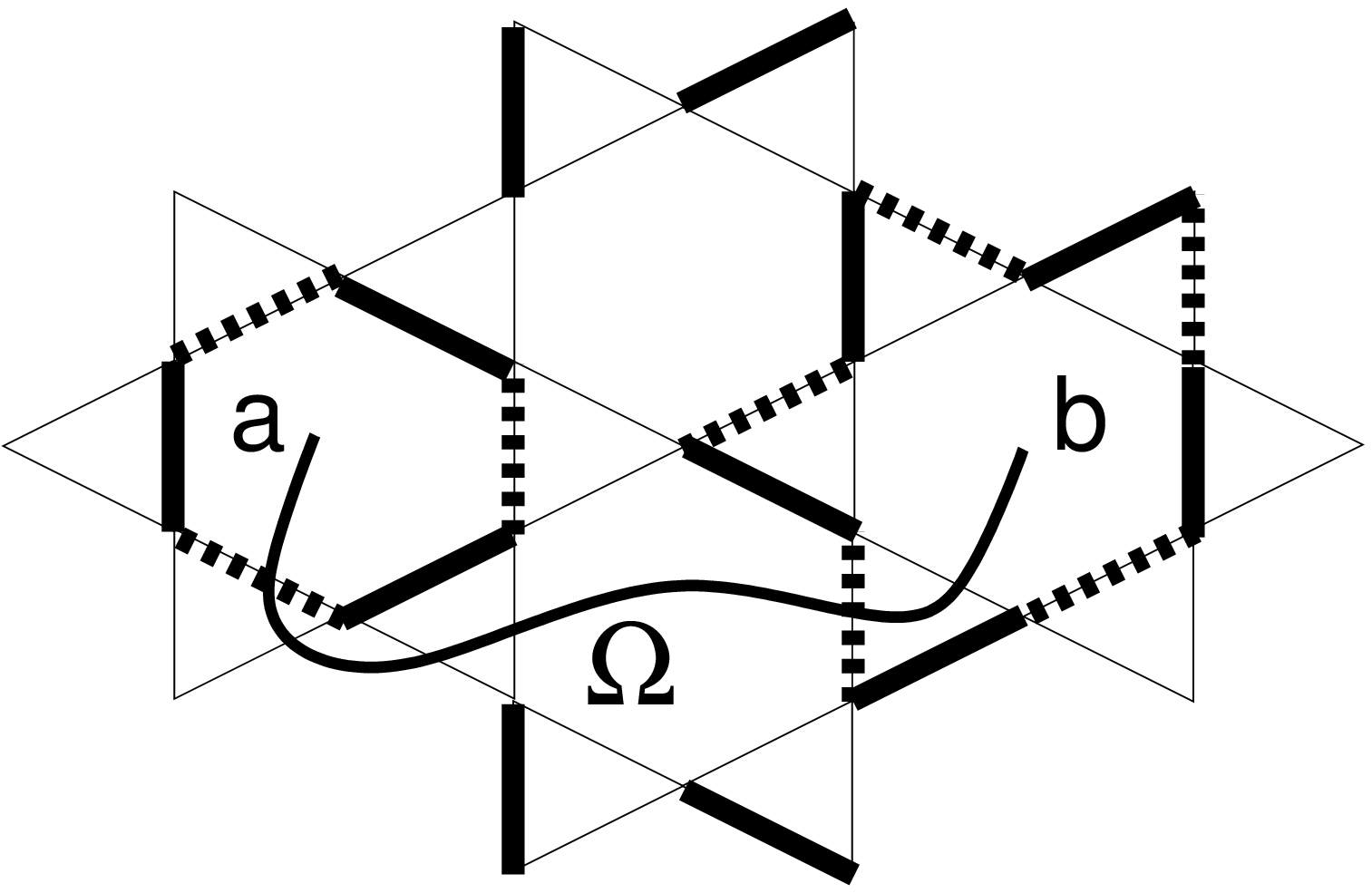}}
	\end{center}
	
	\caption[99]{{\bf a)} A  dimer covering on  the kagome lattice
	(fat  bonds).    The corresponding   arrow  representation  is
	displayed.  The  Kagome lattice is the  medial  lattice of the
	hexagonal  lattice (dashed bonds).  {\bf b)}  A pair of visons
	is  created by  applying  to the   SRK wave-function a  factor
	$(-1)$ for each  dimer crossing the   cut $\Omega$.  It  is an
	exact             eigenstate        of   Eq.~\ref{eq:Hsigmax}.
	}\label{KagHexVison}
\end{figure}

{\em Medial  lattice  construction}.--- The dimer  models described in
this Letter can be defined on lattices made of corner-sharing triangle,
constructed  in  the following way.  Let  $H$  be  a trivalent lattice
(each site has  three neighbors).  The  lattice $K$,  where the dimers
live, is the medial lattice  of $H$, {\it  i.e.}  the sites of $K$ are
the midpoints of the bonds of  $H$ (Fig.~\ref{medial}).  If $H$ is the
honeycomb   lattice,        $K$      is    the      kagome     lattice
(Fig.~\ref{KagHexVison}a).    In   the   following, unless   mentioned
otherwise, we use kagome  for simplicity, where  plaquettes of $H$ are
hexagons.   We  stress however  that  all  results  can be generalized
straightforwardly  to other lattices (squagome~\cite{sg01} lattice for
instance, as well as one-dimensional examples).

{\em Pseudospin representation}.--- Let us begin with the definition of
a simple dimer  model.  For each hexagon  $h$, we  define  an operator
$\sigma^x(h)$  as the   sum  of  all   possible kinetic  energy  terms
involving $h$ only:
\begin{equation}
	\sigma^x(h)=\sum_{\alpha=1}^{32}
	\left|d_\alpha(h) \right>\left< \bar{d}_\alpha(h)\right|
	+
	\left| \bar{d}_\alpha(h)\right>\left<d_\alpha(h)\right|
	\label{eq:sigmax}
\end{equation}
The sum runs over  the  $32$ loops  on  kagome which enclose  a single
hexagon and around which dimers  can be moved (see Ref~\cite{ze95} for
an explicit list).   The  shortest loop is    the hexagon itself,   it
involves 3  dimers.  4, 5  and  6-dimers  moves are  also possible  by
including 2, 4  and 6 additional  triangles  (the loop  length must be
even).  The largest   loop is the star.   For  each loop  $\alpha$  we
associate  the   two ways  dimers    can be placed    along that loop:
$\left|d_\alpha(h)\right>$ and $\left|\bar{d}_\alpha(h)\right>$.

For a given dimer covering $\left|D\right>$, all the kinetic operators
in  the sum but  one  annihilate $\left|D\right>$.  As a  consequence,
$\sigma^x(h)^2=1$.  One can  further check  that these operators  flip
the  pseudospin    variables $\sigma^z(h)$  introduced  by   Elser and
Zeng~\cite{ez93,ze95}  (EZ)  to  label    dimer coverings~\footnote{EZ
pseudospins variables    are different from the  well-known   mapping
[M.~E.~Fisher, J. Math. Phys. {\bf 7},  1776 (1966)] used to solve the
Ising   model on a  planar  lattices  in terms   of  dimers.}.  It  is
important to note that $\sigma^z$ operators depend on  the choice of a
reference $\re$  and     are not  local,   unlike   $\sigma^x$.    The
$\sigma^x(h)$ {\em commute with each other}.  This is not obvious from
Eq.~\ref{eq:sigmax} and it is most easily demonstrated in terms of the
arrow representation that we introduce below.

{\em   Arrow   representation}.---  A   correspondence  between  dimer
coverings on   the  kagome  lattice and {\em     sets  or arrows}   as
illustrated in   Fig.~\ref{KagHexVison}a was  introduced  by Elser and
Zeng~\cite{ez93}.  Each arrow  has two possible directions:  it points
toward the interior of one of the  two neighboring triangles.  If site
$i$ belongs  to a   dimer $(i,j)$ its   arrow  must point   toward the
triangle  the site $j$ belongs   to.  Consider a triangle without  any
dimer,  this   arrow rule implies that   it  will have  three outgoing
arrows.   Other  triangles   will  have  two incoming  arrows  and one
outgoing arrow.   In other words,  the  number of  outgoing arrows  is
constrained to be odd.

The number of dimer  coverings is $2^{N/3+1}$ where  $N$ is the number
of sites~\cite{hw88elser89,ez93}.  $K$ has  $N$ sites and $N$  arrows,
$2N/3$  triangles and  one   constraint per  triangle.   However  only
$2N/3-1$ constraints are  independent  because their product  for  all
triangles is equal to one.  We are left with  $N/3+1$ Ising degrees of
freedom.   The existence of   this arrow representation  is a  central
reason for  which  QDM considerably simplify  on  these lattices.  For
example   $\sigma^x(h)$   translates    very simply   in    the  arrow
representation: it flips the  six  arrows sitting around hexagon  $h$,
which clearly conserves the  constraint for all triangles and commutes
from hexagon to hexagon.

{\em Rokhsar-Kivelson point}.--- Consider the following Hamiltonian:
\begin{equation}
	\mathcal{H}_0=-\Gamma\sum_{h} \sigma^x(h)
	\label{eq:Hsigmax}
\end{equation}
where  the sum  runs over  hexagons  (sites   of the dual   of $H$  in
general).    Although very simple  in   the pseudospin variables, this
Hamiltonian  is not  obviously     solvable when written    with dimer
operators.  In the pseudospin variables  the  ground-state is a  fully
polarized ferromagnet in the  $x$ direction, which  is the sum of  all
pseudospin configurations in the EZ $\sigma^z$ basis.  Back to dimers,
this  is nothing but  the sum of  all dimer configurations  in a given
topological sector, that    is a SRK  wave-function~\cite{rk88}.   The
ground-state  appears to be unique  in  each topological sector, which
gives a global 4-fold degeneracy on the torus.

{\em Correlations}.---  Correlations in  the SRK  state of  the kagome
lattice are particularly  simple: irreducible dimer-dimer correlations
are strictly  zero  when their  corresponding triangles  do not touch.
The arrows on two bonds are independent provided they are not involved
in a common constraint, that is a common triangle.  As a result, dimer
on the  kagome  lattice are the most    possible disordered: they  are
independent above a finite distance.

{\em  Gap}.--- The whole  spectrum is known.  The $\sigma^x$ operators
commute from hexagon to hexagon but physical dimer states must satisfy
$\prod_h\sigma^x(h)=1$.   This   constraint  comes    from the   arrow
representation  since $\prod_h\sigma^x(h)$ flips   all the arrows {\em
twice} and therefore  keeps  all dimerizations unchanged.   The  first
excited state appears not to be  a single but a  {\em pair} of flipped
hexagons with energy cost $\Delta=4\Gamma$.

{\em Visons}.---  Despite   of the  simplicity  of  the model  in  the
pseudospin variables, its excited  states are not local when expressed
with dimer degrees of freedom.  A $\sigma^x(h)=-1$ hexagon is a vortex
excitation (also  called  {\em vison}~\cite{sf01}).  Consider a string
which   goes  from    an  hexagon  $a$  to    and   hexagon  $b$  (see
Fig.~\ref{KagHexVison}b) and  let $\Omega(a,b)$ be the  operator which
measures the parity  $\pm1$ of  the  number  of dimers  crossing  that
string.  $\Omega(a,b)$ commutes with all $\sigma^x(h)$, except for the
ends                  of                 the                   string:
$\sigma^x(a)\Omega(a,b)=-\Omega(a,b)\sigma^x(a)$. A dimer move changes
the sign of  $\Omega(a,b)$ if and only if  the associated loop crosses
the   string an  odd number   of  times, which  can   only  be done by
surrounding  one end of  the   string.  This shows that  $\Omega(a,b)$
flips the $x$ component of  the pseudospin in $a$ and $b$~\footnote{Up
to   a global sign   (reference dependent)  $\Omega(a,b)$  is equal to
$\sigma^z(a)\sigma^z(b)$.}    and   $\Omega(a,b)\left|\psi\right>$  is
precisely  the excited  state  of  energy $4\Gamma$  discussed  above.
Visons appear to be perfectly localized in this model.

{\em Spinons}.--- One can consider  the system  with two {\em  static}
unpaired sites  (spinons or   holes).  As for   others  QDM at  a  SRK
point~\cite{rk88,ms01}, the sum of all dimerizations in a given sector
remains an exact eigenstate.   The energy turns  out to be independent
of the relative distance between spinons, which is a strong indication
that   spinons   would be  also  deconfined     if they  had   kinetic
energy. Eventually, we  note that taking a  spinon around a vison will
change the wave function by a $-1$ factor, as expected~\cite{rc89k89}.

{\em Visons and  topology}.--- On a closed surface  of genus $g$,  the
spectrum has a degeneracy given by the  number of topological sectors
$2^{\rm 2g}$ and  excitations are {\em pairs}   of visons, as  already
mentioned.     When   the    sample  has     edges,   the   constraint
$\prod_h\sigma^x(h)=1$ is not  valid anymore.  To  handle this case we
introduce $\sigma^x(\tilde{h})$ which  flips the arrows along the edge
and   which  restores $\sigma^x(\tilde{h})\prod_h\sigma^x(h)=1$.   The
excitations are still pairs of visons but  one vison can be located in
the hole~\cite{sf01}. In this case the gap reduces to $2\Gamma$. In an
cylinder geometry, as in Ref.~\cite{ioffe02} for instance, we have two
sectors and  a doubly degenerate  spectrum.  It is interesting to note
that this dimer liquid has no low-energy edge states.

{\em Liquid - solid transition}.---
We  consider   a    new    QDM    which   is  a    generalization   of
Eq.~\ref{eq:Hsigmax}:
\begin{equation}	
	\mathcal{H}_1=\mathcal{H}_0%-\Gamma \sum_h \sigma^x(h)
	- J \sum_{\left<h,h'\right>}\sigma^z(h)\sigma^z(h')
	\label{eq:ising}
\end{equation}
where the  second sum runs over  pairs  of neighboring  hexagons.  The
$\sigma^z$   operators  are those defined  by  EZ,  they depend on the
choice       of  a    reference       dimerization  $\re$.    A   term
$\sigma^z(h)\sigma^z(h')$ is  local, it is  $=1$ if the arrow which is
in between $h$  and $h'$ is  in the same  position as in the reference
state  and  $-1$   otherwise.   Since   $\sigma^z(h)\sigma^z(h')$   is
equivalent to   $\Omega(h,h')$,  we see  that such  a  term allows  to
create, annihilate and move visons  in the system.  From another point
of view,  the $J$  term in  Eq.~\ref{eq:ising}  counts the   number of
arrows to  be flipped to recover the  reference state.  If $J$ goes to
$+\infty$, this Hamiltonian obviously  selects the reference  state as
the ground-state and dimers are completely  frozen.  In the pseudospin
language $\mathcal{H}_1$ is an Ising  ferromagnet in transverse field,
which displays a second order phase  transition at a critical value of
$\Gamma$  separating  two   phases:     a  ferromagnetic phase    with
$\left<\sigma^z\right>>0$    and     paramagnetic       phase     with
$\left<\sigma^z\right>=0$.  From what we   know of the $\Gamma=0$  and
$\Gamma\to+\infty$ limits  we can identify the  first one with a dimer
solid and  the second one with  the liquid.  From  the Ising  point of
view the  solid phase  is characterized  by $\left<\sigma^z\right>>0$.
The two   ferromagnetic  Ising states   ($\left<\sigma^z\right>=\pm1$)
correspond  to  the same dimer state  on  a  closed surface but differ
along   the edge   for  open     systems.  The   Ising   magnetization
$\left<\sigma^z(h)\right>$  provides a  non-local order  parameter for
the dimer solid,   $\left<\Omega(h,\infty)\right>$,  which involves  a
string going to infinity.

Up to a sign, $\sigma^z(h)$ is the sum  of a creation and annihilation
operators  of a  vison.    The dimer solidification can   therefore be
interpreted    as the onset   of  off-diagonal  order and  macroscopic
occupation  number (of the zero-momentum  state) for the visons.  This
QDM     realizes    a      condensation     of  topological    defects
(``kinks''~\cite{fs78,kogut} or visons~\cite{sf01}) at the confinement
transition.

Notice that although we call it a  solid, the large-$J$ phase does not
go  with a {\em spatial}  spontaneous symmetry breaking  since the $J$
part of  $\mathcal{H}_1$    depends on an   arbitrary  reference state
through the $\sigma^z$  operators;   this  term acts as  an   external
potential which tends to pin the dimers along the reference state.

{\em $\Z$ gauge  theory}.---  The model  of Eq.~\ref{eq:ising}  is the
dual  of  a   $\Z$  gauge  theory~\cite{wegner}   (in  its Hamiltonian
formulation) where  the gauge degrees of   freedom $\tau^z(i)$ live on
the bonds of $H$ ({\it i.e.} sites of $K$).  By definition $\tau^z(i)$
is the operator which flips the arrow at  site $i$ and gauge-invariant
observable  are     made of  products   of    $\tau^z$  around  closed
loops. Spatial gauge  transformations require the $x$ component, which
we define with respect to the  reference state $\re$: $\tau^x(i)=1$ if
the arrow $i$ has the same orientation as  in $\re$ and $\tau^x(i)=-1$
otherwise.  For every  site of    $H$  (every triangle  of $K$)    the
constraint   reads $\tau^x(i_1)\tau^x(i_2)\tau^x(i_3)=1$ where  $i_1$,
$i_2$ and $i_3$ are the  bonds of $H$ emanating  from that site.  This
expresses the fact that physical states must be gauge invariant.  This
shows a one-to-one correspondence between  physical state of the gauge
theory  and dimer coverings  of $K$  and the  redundancy in the  gauge
theory  is solved   by  the  dimer  coverings.   We  wish   to express
$\mathcal{H}_0$ with the gauge degrees of freedom. Since $\sigma^x(h)$
operator flips  all the  arrows around  $h$, it  becomes the plaquette
operator:
\begin{equation}
	\sigma^x(h)=\prod_{i} \tau^z(i)
	\label{eq:tauz}
\end{equation}
where the product runs over the bonds of  $H$ surrounding $h$.
The Ising interactions can be written
\begin{equation}
	\sigma^z(h)\sigma^z(h')=\tau^x(i)
	\label{eq:taux}
\end{equation}
where $i$   is   the  common    link between     $h$ and   $h'$   (see
Fig.~\ref{medial}).  As  a   result   the QDM  of   Eq.~\ref{eq:ising}
translates into the Hamiltonian of a $\Z$  gauge theory (in continuous
time), which is a manifestation of the well-known duality between $\Z$
gauge theories and Ising models~\cite{wegner,kogut}.
Such  a $\Z$ gauge  theory  has a  confined  and deconfined phases.  A
classical  result  is that   they  can be  distinguished through   the
expectation value of a gauge-invariant Wilson loop~\cite{wegner,kogut}
\begin{equation}
	W(\omega)=< \prod_{i\in \partial\omega} \tau^z(i) >
	=
	< \prod_{i\in \omega} \sigma^x(i) >
	\label{eq:W}
\end{equation}
where the first product runs over a  close loop $\partial\omega$ which
surrounds  the area  $\omega$.    $W(\omega)$  changes  from a    {\em
perimeter law}  $\sim\exp(-|\partial\omega|)$   to an {\em   area law}
$\sim\exp(-|\omega|)$ when going from the deconfined at large $\Gamma$
to   the  confined  phase  at large  $J$.      The right-hand  side of
Eq.~\ref{eq:W} expresses $W(\omega)$ as a correlation function for the
dimer   problem.   $W(\omega)$   moves    dimers    along  the    loop
$\partial\omega$ and the mapping to the gauge theory tells us that its
long-distance behavior characterizes the liquid and frozen phases.

{\em  Liquid   -   crystal transition}.---  $\mathcal{H}_1$  has  the
disadvantage of  depending on a  specific reference state.   To remedy
for that we introduce  a QDM which  is reference free and restores the
full lattice symmetry:
\begin{equation}	
	\mathcal{H}_2=\mathcal{H}_0
	- J \sum_{h}s^z(h)
	\label{eq:h2}
\end{equation}
where $s^z(h)=\pm1$ is diagonal in the dimer basis and counts a factor
$-1$  per  anticlockwise arrow   around   the  hexagon $h$.    In  the
$J\to\infty$ limit the system selects $s^z(h)=1$ everywhere.  This can
be   achieved  if the   system spontaneously   breaks  the translation
symmetry and crystallizes in an ordered pattern of six-dimer ``stars''
described in Ref.~\cite{sm02}.   These  ordered states are  degenerate
with  others ($\sim2^{3L}$ where $L$ is  the  linear size) obtained by
shifting the dimers along any straight line.

We have looked numerically (diagonalizations up to 54 kagome sites) at
$\mathcal{H}_2$ and  found  evidence   for a  single   and  continuous
transition from   the      liquid   to    the  star  crystal        at
$J\simeq\Gamma$. This appears as a collapse of the first excitation of
the  liquid, which then transforms  into  a degenerate ground-state of
the crystal phase.   We claim that the critical   point is exactly  at
$J=\Gamma$ from a duality  argument.  From the arrow representation it
is  clear that $\sigma^x(h)$ and $s^z(h')$  commute except  if $h$ and
$h'$  are neighbors,  in  which  case  they  anticommute.   Indeed the
algebraic relations of  $\sigma^x$ and $s^z$ are completely symmetric.
In particular, we observed numerically that the ground-state energy is
{\em exactly}  symmetric with respect  to the exchange of $\Gamma$ and
$J$ (this is not true for all the  eigenstates) and the critical point
must lie at the self-dual point $J=\Gamma$.

{\em Conclusions}.--- The  kagome lattice has the  remarkable property
that dimer  coverings correspond to  the physical  states of  an Ising
model  on  the triangular  lattice  and, by duality,  of  a $\Z$ gauge
theory on the hexagonal one.  Exploiting  this, we introduced QDMs for
which exact results  are derived. In particular,  we obtained for  the
first time the full spectrum of a  QDM realizing a dimer liquid phase.
It  explicitly  realizes fractionalized  excitations  and  topological
order.  Through   several models we   showed that  QDM  on the  kagome
lattice  are    very simple and  natural     tools to investigate  the
connexions between  frustrated  magnets, RVB  physics  and spin-charge
separation.

{\em Acknowledgments}.--- We   are grateful to M.~Gaudin,  K.~Mallick,
R.~Moessner,  C.~Lhuillier  and  M.~Feigel'man  for  several  fruitful
discussions. Numerical diagonalizations of QDM models were done on the
Compacq alpha server of the CEA under project 550.

%___________________________________________________________________

\end{document}